# Refinement of Pipe-and-Filter Architectures

Jan Philipps and Bernhard Rumpe

Institut für Informatik,
Technische Universität München,
D-80290 München,
`www4.in.tum.de/~{philipps,rumpe}`

**Abstract.** Software and hardware architectures are prone to modifications. We demonstrate how a mathematically founded powerful refinement calculus for a class of architectures, namely pipe and filter architectures, can be used to modify a system in a provably correct way. The calculus consists of basic rules to add and to remove filters (components) and pipes (channels) to a system. A networking example demonstrates some of the features of our calculus.
The calculus is simple, flexible and compositional. Thus it allows us to build more complex and specific rules that e.g. embed models of existing architectures or define design patterns as transformation rules.

## 1 Introduction

Only in recent years, fueled through the survey [10] and greatly enhanced by [16], notations to define software architectures have been developed (see [12] for an overview). One specific kind of architecture description style — the pipe and filter approach —, is especially useful for asynchronously communicating, distributed processes.

The architecture of a software or hardware system influences its efficiency, its adaptability, and the reusability of software components. Especially the adaptation to new requirements causes frequent changes in the architecture while the system is developed, or when it is later extended. However, the definition of architecture is still rather informal in the software engineering community, and the question of how to properly modify an architecture has not been adequately addressed so far.

In this paper, we examine how pipe and filter achitectures can be modified, so that the new system is a provably correct refinement of the original system. Filters are units of concurrency and computation, connected only through pipes to asynchronously send and receive messages along them. This model corresponds to data flow networks and we therefore will interchangeably use the words "component" and "filter" as well as "channel" and "pipe". In fact, we prefer the term component over filter throughout this paper, as it fits better to the hierarchic nature and the explicitly defined interfaces of these units. Also the the term channel is preferred over pipe. Our work is based on a precise mathematical model [4–6] for data flow networks. An earlier version of this calculus [14] is improved and its powerful features demonstrated in a networking example, where a previously given architecture is optimized. Our model gives a compositional semantics to data flow networks (pipe and filter architectures with feedback), and hence components can be composed to build hierarchical models of a system. Thus, our semantic model does not only contain pipe and filters but also asynchronous communication processes and event based systems, all three listed as different architectural styles in [16].

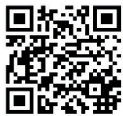

The semantic model is simple, yet powerful: when specifying a component behavior, certain aspects can be left open. We refer to this style as *underspecification*. The reduction of this underspecification immediately gives a refinement relation for blackbox behaviors of components.

In addition to blackbox or *behavioral* refinement, two other classes of refinement relations are established today, namely *structural refinement* (glassbox refinement) and *signature refinement*. While blackbox refinement only relates blackbox behaviors of not further detailed components, structural refinement allows us to refine a blackbox behavior by a subsystem architecture. Signature refinement deals with the manipulation of the system or component interfaces. As shown in [2], both structural and signature refinement can be reduced to behavioral refinement. In Section 2 we will see that behavioral refinement is a simple subset relation.

Neither of these three refinement classes, however, allows architectural refinement in the sense that two glassbox architectures are related. In [14], we introduced a concept for glassbox refinement; again, it can be defined in terms of behavioral refinement. For the practical application of architectural refinement, we defined a rule system to incrementally change an architecture, e.g. by adding new components or channels.

In this paper, we demonstrate in detail how the rule system can be applied to a concrete example. It is structured as follows. In Sections 2 and 3 we present the mathematical foundations and define the concepts of component and system. In Section 4 we summarize the rules introduced in [14]. Section 5 describes the refinement of a simple data acquisition system. Section 6 concludes.

## 2 Semantic Model

In this section we introduce the basic mathematical concepts for the description of systems. We concentrate on interactive systems that communicate asynchronously through channels. A component is modeled as a relation over input and output communication histories that obeys certain causality constraints. We assume that there is a given set of channel identifiers, $\mathbb{C}$, and a given set of messages, $M$.

*Streams.* We use *streams* to describe communication histories on channels. A stream over the set $M$ is a finite or infinite sequence of elements from $M$. By $M^*$ we denote the finite sequences over the set $M$. The set $M^*$ includes the empty sequence that we write as $\langle\rangle$. The set of infinite sequences over $M$ is denoted by $M^\infty$.

Communication histories are represented by *timed streams*: $M^\aleph =_{def} (M^*)^\infty$
The intuition is that the time axis is divided into an infinite stream of time
intervals, where in each interval a finite number of messages may be transmitted.
In Fig. 1, a prefix of such a stream is visualized. It is based on the messages
$\{a, b, c\}$ which occur several times during the first six units of time.

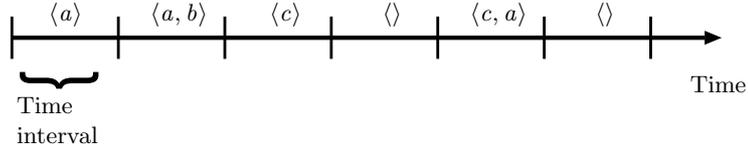

**Fig. 1.** Prefix of a communication history

These intervals can be considered to be of a fixed duration, such as months
or days for reports in business information systems, or milliseconds in more
technical applications. Their duration need not be fixed, however: the intervals
could also span the time between certain events that are of interest to the system,
such as the pressing of a button. In each interval, the order of the messages is
fixed, but the exact arrival time of a message is unknown.

For $i \in \mathbb{N}$ and $x \in M^\aleph$ we denote by $x{\downarrow}i$ the sequence of the first $i$ sequences
in the stream $x$. When writing specifications, we sometimes abstract from the
interval boundaries, and regard a stream as the finite or infinite sequence of
messages that results from the concatenation of all the intervals. We then use
the syntax $a \,\&\, r$ to split a stream into its first element $a$, and the remaining
sequence $r$.

A *named stream tuple* is a function $\mathbb{C} \to M^\aleph$ that assigns histories to channel
names. For $C \subseteq \mathbb{C}$ we write $\vec{C}$ for the set of named stream tuples with domain
$C$. For $x \in \vec{C}$ and $C' \subseteq C$, the named stream tuple $x \mid_{C'} \in \vec{C'}$ denotes the
restriction of $x$ to the channels in $C'$:

$$\forall\, c \in C' : x \mid_{C'} (c) = x(c)$$

We extend operators like $\downarrow$ and $\mid$ in the usual pointwise style to stream tuples
and sets over stream tuples.

*Behaviors.* We model the interface behavior of a component with the set of input
channels $I \subseteq \mathbb{C}$ and the set of output channels $O \subseteq \mathbb{C}$ by a function

$$\beta : \vec{I} \to \mathbb{P}(\vec{O})$$

Intuitively, $\beta$ maps the incoming input on $I$ to the set of possible outputs on
$O$, and thus describes the visible behavior of a component with input channels
$I$ and output channels $O$. Equivalently, $\beta$ can be seen as a relation over the
named stream tuples in $\vec{I}$ and the named stream tuples in $\vec{O}$. $\beta$ is called a

*behavior*. Since for every input history multiple output histories can be allowed by a behavior, it is possible to model nondeterminism, or equivalently, to regard relations with multiple outputs for one input as underspecified.

A function $f \in \vec{I} \to \vec{O}$ can be seen as a special case of a deterministic relation. We say the $f$ is *time guarded*, iff for all input histories $x$ and $y$, and for all $i \in \mathbb{N}$

$$x\downarrow i = y\downarrow i \Rightarrow (f\ x)\downarrow(i+1) = (f\ y)\downarrow(i+1)$$

A time guarded function $f$ is called a *strategy* for a behavior $\beta$ if for all $x$ we have $f(x) \in \beta(x)$. If $\beta$ has at least one strategy, we say that $\beta$ is *realizable*.

Time guardedness reflects the notion of time and causality. The output at a certain time interval may only depend on the input received so far, and not on future input.

*Interface adaption.* Given a behavior $\beta : \vec{I} \to \mathbb{P}(\vec{O})$, we can define a behavior with a different interface by extending the set of input channels, and restricting the set of output channels. If $I \subseteq I'$ and $O' \subseteq O$, then $\beta' = \beta\updownarrow_{O'}^{I'}$ is again a behavior with $\beta'(i) = (\beta(i\mid_I))\mid_{O'}$.

This corresponds to the change of the component interface by adding input channels that are ignored by the component, and by removing output channels that are ignored by the environment.

*Composition.* Behaviors can be composed by a variety of operators. Sequential and parallel composition, as well as a feedback construction is introduced in [11]. For our work, we use a generalized operator $\otimes$ that composes a finite set of behaviors

$$B = \{\beta_1 : \vec{I_1} \to \mathbb{P}(\vec{O_1}), \ldots, \beta_n : \vec{I_n} \to \mathbb{P}(\vec{O_n})\}$$

in parallel with implicit feedback. We define

$$O = \cup_{1 \leq k \leq n} O_k$$
$$I = (\cup_{1 \leq k \leq n} I_k) \setminus O$$

where $O$ is the union of all component outputs, and $I$ is the set of those inputs, that are not connected to any of the components' outputs. Then the relation $\otimes B \in \vec{I} \to \mathbb{P}(\vec{O})$ is characterized by:

$$o \in (\otimes B)(i) \Leftrightarrow$$
$$\exists\, l \in \overrightarrow{(I \cup O)}:$$
$$l\mid_O = o \wedge l\mid_I = i \wedge$$
$$\forall\, k \in \{1, \ldots n\} :\ l\mid_{O_k} \in \beta_k(l\mid_{I_k})$$

If all behaviors in $B$ are realizable, then so is $\otimes B$. The proof follows [11]; it relies on the time guardedness of strategy functions. It is easy to express parallel and sequential composition of behaviors with the $\otimes$ operator.

*Refinement.* Intuitively, a behavior describes the externally observable relation between the input and the output that the clients of a component may rely on. Refining a behavior in a modular way means that the client's demands are still met, when the component behavior is specialized. Formally, the refinement relation in our framework is defined as follows. Given two behaviors $\beta_1, \beta_2 \in \vec{I} \to \mathbb{P}(\vec{O})$ we say that $\beta_1$ is refined by $\beta_2$, iff

$$\forall\, i \in \vec{I} : \beta_2(i) \subseteq \beta_1(i)$$

Refinement means in our context that each possible channel history of the new component is also a possible channel history of the original component.

## 3   Components and Systems

In this section, we define an abstract notion of system architecture. Basically, a system consists of a set of *components* and their *connections*. We first define components, and then introduce the architectural or glassbox view, and the blackbox view of a system.

*Components.* A *component* is a tuple $c = (n, I, O, \beta)$, where $n$ is the name of the component, $I \subseteq \mathbb{C}$ is the set of input channels, and $O \subseteq \mathbb{C}$ the set of output channels. Moreover, $\beta : \vec{I} \to \mathbb{P}(\vec{O})$ is a behavior. The operators name.$c$, in.$c$, out.$c$ and behav.$c$ yield $n$, $I$, $O$ and $\beta$, respectively. The name $n$ is introduced mainly as a convenience for the system designer. The channel identifiers in.$c$ and out.$c$ define the interface of the component. This interface can be visualized as shown in Fig. 2.

*Architectural view of a system.* In the architectural view, a system comprises a finite set of components. A connection between components is established by using the same channel name.

A system is thus a tuple $S = (I, O, C)$, where $I \subseteq \mathbb{C}$ is the input interface, and $O \subseteq \mathbb{C}$ is the output interface of the system. $C$ is a finite set of components. Like single components, a system can be visualized, describing the structure of the connections between its components, as given in Fig. 3.

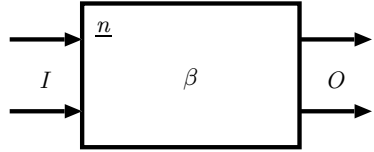

**Fig. 2.** Component diagram for $(n, I, O, \beta)$

We want to be able to decompose systems hierarchically. In fact, as we will see, a system can be regarded as an ordinary component. Therefore systems need

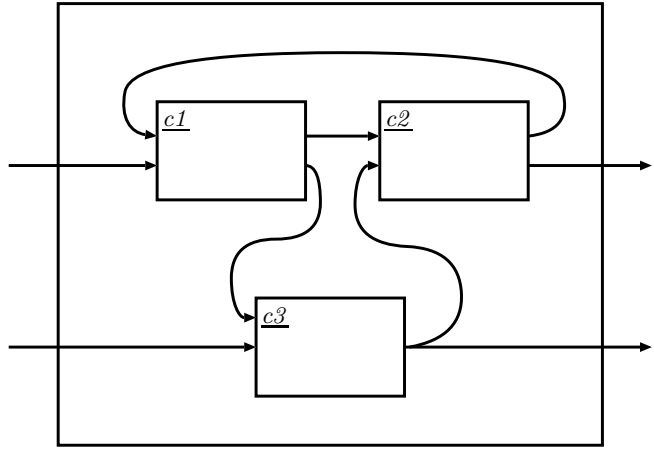

**Fig. 3.** System structure diagram

not be closed, and we introduce the interface channels $I$ and $O$ to distinguish external from internal channels. We define the operators in.$S$, out.$S$, arch.$S$ to return $I$, $O$ and $C$, respectively. In addition, we write:

in.$C =_{def} \cup_{c \in C} (\text{in}.c)$
out.$C =_{def} \cup_{c \in C} (\text{out}.c)$

for the union of the input or output interfaces, respectively, of the components of $S$.

The following consistency conditions ensure a meaningful architectural view of a system $S$. For all components $c_1, c_2 \in \text{arch}.S$ with $c_1 \neq c_2$, we require the following properties:

1. Different components have different names: name.$c_1 \neq$ name.$c_2$
2. Each channel is controlled by only one component: out.$c_1 \cap$ out.$c_2 = \varnothing$
3. Input channels of the system interface are controlled by the environment, not by a component: in.$S \cap$ out.$C = \varnothing$
4. Each input channel of a component is controlled by either a component or by the environment: in.$C \subseteq$ out.$C \cup$ in.$S$
5. Each channel of the output interface is controlled by a component: out.$S \subseteq$ out.$C$.

Note that we allow that input channels are in more than one interface: a channel can have multiple readers, even broadcasting is possible. Not every channel of the system input interface has to be connected to a component, since condition 4 only requires the subset relation instead of equality. We allow a component to read and write on the same channel if desired; as a consequence of conditions (3) and (5), however, input and output of a composed system are disjoint.

*Blackbox view of a system.* The behavior of a component $c$ is given in terms of its relation $\mathsf{behav}.c$ between input and output streams. We define the *blackbox behavior* of a system $S$ composed of finitely many components $\mathsf{arch}.S$ using the composition operator $\otimes$. The result of this composition is then made compatible with the system interface by restricting the output channels to those in $\mathsf{out}.S$, and by extending the input channels to those in $\mathsf{in}.S$:

$$[\![S]\!] = (\otimes\{\ \mathsf{behav}.c\ \mid\ c \in \mathsf{arch}.S\ \})\updownarrow_{\mathsf{out}.S}^{\mathsf{in}.S}$$

Because of the context conditions for systems the composition is well-defined. The hiding of the internal output channels $\mathsf{out}.C \setminus \mathsf{out}.S$ and the extension with the unused input channels $\mathsf{in}.S \setminus \mathsf{in}.C$ is also well-defined.

The blackbox behavior has the signature:

$$[\![S]\!] : \overrightarrow{\mathsf{in}.S} \to \mathbb{P}(\overrightarrow{\mathsf{out}.S})$$

Thus, the blackbox behavior can now be used as a component description itself. Introducing a fresh name $n$, we define the component $c_S$ as:

$$c_S = (n, \mathsf{in}.S, \mathsf{out}.S, [\![S]\!])$$

In this way, a hierarchy of architectural views can be defined and iteratively refined and detailed. Later on we need a more detailed definition of this semantics. By expanding the definitions of the $\otimes$ and $\updownarrow$ operators, we obtain the following equivalent characterisation of $[\![(I, O, C)]\!]$:

$$\begin{aligned}
&o \in [\![(I, O, C)]\!](i) \Leftrightarrow \\
&\quad \exists\, l \in \overrightarrow{(I \cup \mathsf{out}.C)} : \\
&\quad\quad l\mid_O = o \land l\mid_I = i \land \\
&\quad\quad\quad \forall\, c \in C :\ l\mid_{\mathsf{out}.c} \in (\mathsf{behav}.c)(l\mid_{\mathsf{in}.c})
\end{aligned}$$

This expanded characterisation says, that $o$ is an output of the system for input $i$ (line 1), iff there is a mapping $l$ of all channels to streams (line 2), such that $l$ coincides with the given input $i$ and output $o$ on the system interface channels (line 3) and feeding the proper submapping of $l$ into a component results also in a submapping of $l$.

## 4  Refinement of Architectures

When a system architecture is refined, it must not break the interaction with its environment. The observable behavior of a refined system architecture must be a refinement of the behavior of the original system architecture. In this paper, we leave the interface of the system architecture unchanged. Interface refinements that affect the signature of a system $S$ are described in [2] for blackbox behaviors; they can be adapted to our architectural framework. We also ignore aspects of realizability. The techniques used to prove that a component specification is realizable are orthogonal to the rules presented here, and will not be considered in this paper. We therefore define the refinement relation on system architectures as a behavioral refinement on the given interface:

$$S \leadsto S' \Leftrightarrow_{def} \forall\, i \in \overrightarrow{\mathsf{in}.S} : [\![S']\!](i) \subseteq [\![S]\!](i)$$

As explained above, we tacitly assume that $\mathsf{in}.S = \mathsf{in}.S'$ and $\mathsf{out}.S = \mathsf{out}.S'$. Stepwise refinement is possible, since the refinement relation is transitive:

$$S \leadsto S' \wedge S' \leadsto S'' \Rightarrow S \leadsto S''$$

In [14], we defined and justified a set of constructive refinement rules that allows refinements of system architectures. The rules allow us to add and remove components, to add and remove channels, to refine the behavior of components, and to refine single components to subsystems and vice versa. In the sequel, we summarize these rules; in Section 5 we apply them to a simple example. Each rule refines a system architecture $S = (I, O, C)$ into another system architecture $S' = (I, O, C')$. We use the syntax

$$S \text{ WITH } C := C'$$

to denote the architecture $(I, O, C')$. In addition, we write

$$S \text{ WITH } c := c'$$

to denote the architecture $(I, O, (C \setminus \{c\}) \cup \{c'\})$. To create a component with the same name and interface as $c = (n, I, O, \beta)$, but with a different behavior $\beta'$, we use the syntax

$$c \text{ WITH } \mathsf{behav}.c := \beta'$$

to denote the component $(n, I, O, \beta')$. Similarly, we can change the name or interface of a component. The refinement rules are presented in the syntax

$$\frac{(Premises)}{(Refinement)}$$

where the premises are conditions to be fulfilled for the refinement relation to hold.

*Behavioral refinement.* Systems can be refined by refining the behavior of their components. Let $c \in C$ be a component. If we refine the behavior of $c$ to $\beta$, we get a refinement of the externally visible, global system behavior:

$$\frac{c \in C \qquad \forall\, i \in \overrightarrow{\mathsf{in}.c} : \ \beta(i) \subseteq \mathsf{behav}.c(i)}{S \leadsto S \text{ WITH } \mathsf{behav}.c := \beta}$$

In some cases, to prove the behavioral refinement of $c$ some assumptions on the contents of $c$'s input channels are necessary. Then this simple rule cannot be used.

To overcome this problem, we introduce the notion of behavioral refinement in the context of an *invariant*. An invariant is a predicate $\Psi$ over the possible message flows within an architecture $S = (I, O, C)$:

$$\Psi : \overrightarrow{(I \cup \mathsf{out}.C)} \to \mathbb{B}$$

An invariant is valid within an architecture, if it holds for all named stream tuples $l$ defining the system's streams. This can be formally expressed similar to the expanded definition of the system semantics $[\![S]\!]$ presented in Section 3:

$$\forall\, l \in \overrightarrow{(I \cup \mathsf{out}.C)} :$$
$$\quad (\forall\, c \in C : l \mid_{\mathsf{out}.c} \in (\mathsf{behav}.c)(l \mid_{\mathsf{in}.c})) \Rightarrow \Psi(l)$$

Note that invariants are not allowed to restrict the possible inputs on channels from $I$, but only characterize the internal message flow.

Let us now assume that we want to replace the behavior of component $c$ by a new behavior $\beta$. The latter is a refinement of $\mathsf{behav}.c$ under the invariant $\Psi$, when:

$$\forall\, l \in \overrightarrow{(I \cup \mathsf{out}.C)} :$$
$$\quad \Psi(l) \Rightarrow \beta(l \mid_{\mathsf{in}.c}) \subseteq (\mathsf{behav}.c)(l \mid_{\mathsf{in}.c})$$

Thus, the complete refinement rule is as follows. The two premises express that $\Psi$ is a valid invariant, and that $\beta$ refines $\mathsf{behav}.c$ under this invariant.

$$\left|\begin{array}{l} \forall\, l \in \overrightarrow{(I \cup \mathsf{out}.C)} : \\ \quad (\forall\, c \in C : l \mid_{\mathsf{out}.c} \in (\mathsf{behav}.c)(l \mid_{\mathsf{in}.c})) \Rightarrow \Psi(l) \\ \forall\, l \in \overrightarrow{(I \cup \mathsf{out}.C)} : \\ \quad \Psi(l) \Rightarrow \beta(l \mid_{\mathsf{in}.c}) \subseteq (\mathsf{behav}.c)(l \mid_{\mathsf{in}.c}) \\ \hline S \leadsto S \text{ WITH } \mathsf{behav}.c := \beta \end{array}\right.$$

This rule is the only one that requires global properties of a system as a premise. The other rules only deal locally with one affected component. However, since $\Psi$ is used only for a single application of this rule, it is often sufficient to prove its invariance with respect to a relevant subset of all the system components.

Behavioral refinement of a component usually leads to true behavioral refinement of the system architecture. This is in general not the case for the following architectural refinements, which leave the global system behavior unchanged.

*Adding and removing output channels.* If a channel is neither connected to a system component, nor part of the system interface, it may be added as a new output channel to a component $c \in \mathsf{arch}.S$:

$$
\begin{array}{|l}
p \in \mathbb{C} \setminus (I \cup \mathsf{out}.C) \\
\beta \in \overrightarrow{\mathsf{in}.c} \to \mathbb{P}(\overrightarrow{\mathsf{out}.c \cup \{p\}}) \\
\forall i, o : \ o \in \beta(i) \Leftrightarrow o \mid_{\mathsf{out}.c} \in \mathsf{behav}.c(i) \\
\hline
S \rightsquigarrow S \text{ WITH} \\
\qquad \mathsf{out}.c := \mathsf{out}.c \cup \{p\} \\
\qquad \mathsf{behav}.c := \beta
\end{array}
$$

The new behavior $\beta$ does not restrict the possible output on channel $p$. Hence, the introduction of new output channels increases the nondeterminism of the component. It does not, however, affect the behavior of the composed system, since $p$ is neither part of the system interface nor connected to any other component. The contents of the new channel can be restricted with the behavioral refinement rule.

Similarly, an output channel $p \in \mathsf{out}.c$ can be removed from the component $c$, provided that it is not used elsewhere in the system architecture:

$$
\begin{array}{|l}
p \notin O \cup \mathsf{in}.C \\
\beta = \mathsf{behav}.c\Updownarrow_{\mathsf{out}.c \setminus \{p\}}^{\mathsf{in}.c} \\
\hline
S \rightsquigarrow S \text{ WITH} \\
\qquad \mathsf{out}.c := \mathsf{out}.c \setminus \{p\} \\
\qquad \mathsf{behav}.c := \beta
\end{array}
$$

The new behavior $\beta$ is the restriction of the component behavior $\mathsf{behav}.c$ to the remaining channels.

Adding and removing output channels are complementary transformations. Consequently, both rules are behavior preserving. This is not surprising, since the channel in question so far is not used by any other component.

*Adding and removing input channels.* An input channel $p \in \mathbb{C}$ may be added to a component $c \in C$, if it is already connected to the output of some other component or to the input from the environment:

$$
\begin{array}{|l}
p \in I \cup \mathsf{out}.C \\
\beta = \mathsf{behav}.c\Updownarrow_{\mathsf{out}.c}^{\mathsf{in}.c \cup \{p\}} \\
\hline
S \rightsquigarrow S \text{ WITH} \\
\qquad \mathsf{in}.c := \mathsf{in}.c \cup \{p\} \\
\qquad \mathsf{behav}.c := \beta
\end{array}
$$

The behavior $\beta$ now receives input from the new input channel $p$, but is still independent of the data in $p$.

If the behavior of a component $c$ does not depend on the input from a channel $p$, the channel may be removed:

$$
\begin{array}{|l}
\forall\, i, i' \in \overrightarrow{\mathsf{in}.c}:\ i\mid_{\mathsf{in}.c\setminus\{p\}} = i'\mid_{\mathsf{in}.c\setminus\{p\}} \\
\qquad\Rightarrow \mathsf{behav}.c(i) = \mathsf{behav}.c(i') \\
\forall\, i \in \overrightarrow{\mathsf{in}.c}:\ \beta(i\mid_{\mathsf{in}.c\setminus\{p\}}) = \mathsf{behav}.c(i) \\
\hline
S \rightsquigarrow S \text{ WITH} \\
\qquad \mathsf{in}.c := \mathsf{in}.c \setminus \{p\} \\
\qquad \mathsf{behav}.c := \beta
\end{array}
$$

Because the component does not depend on the input from $p$ (first premise), there is a behavior $\beta$ satisfying the second premise. The rule for removing input channels might seem useless — why should a component have an input it does not rely on? However, note that it is possible to first add new input channels that provide basically the same information as an existing channel, then to change the component's behavior so that it relies on the new channels instead. Finally, the old channel can safely be reduced. As with output channels, adding and removing input channels are complementary transformations and thus behavior preserving. This is because the input channels do not influence the component's behavior, and therefore the global system behavior is unchanged, too.

*Adding and removing components.* A component can be added without changing the global system behavior if we ensure that it is not connected to the other components, or to the system environment. Later, we may successively add input or output channels, and refine the new component's behavior with the previously given rules.

$$
\begin{array}{|l}
\forall\, c \in C:\ \mathsf{name}.c \neq n \\
\hline
S \rightsquigarrow S \text{ WITH } C := C \cup \{(n, \varnothing, \varnothing, \alpha)\}
\end{array}
$$

The premise simply ensures that the name $n$ is fresh; the new behavior $\alpha$ is somewhat subtle: it is the unique behavior of a component with no input and no output channels: $\{()\} = \alpha(())$. Similarly, components may be removed if they have no output ports that might influence the functionality of the system.

$$
\begin{array}{|l}
\mathsf{out}.c = \varnothing \\
\hline
S \rightsquigarrow S \text{ WITH } C := C \setminus \{c\}
\end{array}
$$

Note that the addition of a component without input and output channels necessarily does not affect the system behavior. The trivial behavior of this component can made less deterministic by adding channels and more deterministic by refining its behavior.

*Expanding and Folding.* As we have seen, components can be defined with the blackbox view of system architectures. In this way system architectures can be decomposed hierarchically: a single component of a system is replaced by another system. We therefore define a rule for expansion of components. Assume a given system architecture $S = (I_S, O_S, C_S)$ contains a component $c \in C_S$.

This component $c$ is itself described by an architecture $T = (I_T, O_T, C_T)$. The names of the components in $T$ are assumed to be disjoint from those in $S$; through renaming this can always be ensured. The expansion of $T$ in $S$ takes the components and channels of $T$ and incorporates them within $S$.

$$
\begin{array}{|l}
c = (n, I_T, O_T, [\![T]\!]) \\
\mathsf{out}.C_T \cap \mathsf{out}.C_S = \mathsf{out}.c \\
\mathsf{out}.C_T \cap I_S = \varnothing \\
\hline
S \rightsquigarrow S \text{ WITH } C_S := C_S \setminus \{c\} \cup C_T
\end{array}
$$

The first premise means that the architecture $T$ describes the component $c$. The other two premises require that the internal channels of $T$, which are given by $\mathsf{out}.C_T \setminus O_T$, are not used in $S$. In general, this can be accomplished through a renaming rule, which it would be straightforward to define.

The complementary operation to the expansion of a component is the folding of a subarchitecture $T = (I_T, O_T, C_T)$ of a given system $S = (I, O, C)$. This rule is equipped with several context conditions that are mainly concerned with the identification of a correct subarchitecture to be folded.

$T$ is a subarchitecture of $S$, if

- the components $C_T$ are a subset of the components $C$ of $S$;
- the inputs $I_T$ of $T$ contain at least the channels that are needed, but not provided by components of $T$: $\mathsf{in}.C_T \setminus \mathsf{out}.C_T \subseteq I_T$;
- the inputs $I_T$ may contain more input channels, but only those who are provided from outside: $I_T \subseteq I \cup \mathsf{out}.C$,  $I_T \cap \mathsf{out}.C_T = \varnothing$;
- similarly, the outputs $O_T$ are a subset of the component outputs $\mathsf{out}.C_T$: $O_T \subseteq \mathsf{out}.C_T$;
- the outputs $O_T$ include at least those outputs from $\mathsf{out}.C_T$ that are needed by a component outside of $T$: $\mathsf{out}.C_T \cap \mathsf{in}.(C \setminus C_T) \subseteq O_T$;
- and the outputs $O_T$ include those outputs from $\mathsf{out}.C_T$ that are part of the interface of $S$: $\mathsf{out}.C_T \cap O \subseteq O_T$.

The folding rule is then defined as follows, containing a compact description of the above stated requirements for $I_T$ and $O_T$:

$$
\begin{array}{|l}
C_T \subseteq C \\
\mathsf{in}.C_T \setminus \mathsf{out}.C_T \subseteq I_T \subseteq (I \cup \mathsf{out}.C) \setminus \mathsf{out}.C_T \\
\mathsf{out}.C_T \cap (O \cup \mathsf{in}.(C \setminus C_T)) \subseteq O_T \subseteq \mathsf{out}.C_T \\
\forall\, c \in C \setminus C_T : \ \mathsf{name}.c \neq n \\
\hline
S \rightsquigarrow S \text{ WITH } C := C \setminus C_T \cup \{(n, I_T, O_T, [\![T]\!])\}
\end{array}
$$

The first three premises are the conditions mentioned above; the fourth premise requires that the name $n$ of the new component is not used elsewhere in the resulting system. In practice, such a folding is done simply by selecting a subset of the components of $S$ and pushing an appropriate "folding"-button.

Most of the rules presented might seem to be simple. However, as demonstrated in the next section, because of the transitivity of the refinement relation the refinement rules can be composed to complex system modifications. This allows to group rules together to derive complex rules as best-practice rules often used as refinement steps.

## 5 Refinement Example

In this section, we demonstrate how our refinement rule system can be used in practice. Our example architecture is shown in Fig. 4; it models a small data acquisition system. A similar example was carried out in [17].

This example covers a simple, yet frequent architecture modification, which is problematic for the usual functional decomposition techniques. We apply eight refinement steps in this example. Most of them are very simple; they could be carried out by clicking within an appropriate diagram-based tool. No "proof-support" in the stricter sense is necessary. One rule, however, uses a kind of invariant, that resembles both a provision of the changed component as well as an assumption about a part of the environment. We prove this invariant by hand. In practice, the correctness of the refinement could be done by tool support, or just by reviewing the generated proof conditions.

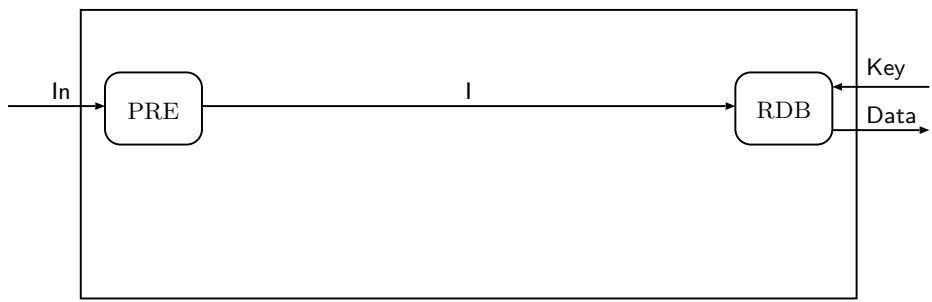

**Fig. 4.** Database example

The system reads input via an input In; the messages on In consist of pairs of a key and some data to be stored under this key; new data values for the same key overwrite old values. Concurrently, the system answers request for the data of a certain key that is input via channel Key by transmitting the data stored in the database under this key via channel Data. Internally, the system consists of two components: a preprocessor PRE, and a database RDB. The data from the environment first undergoes some transformations in PRE, and is then forwarded via the internal channel I to the remote database.

Let $Key$ be the set of possible keys for the database, and $Data$ the set of possible data values. Then, $Entry = Key \times Data$ is the set of possible entries for the database. The database itself is modeled as a function $M : Key \to Data$. We write $M(k)$ for the data item stored under key $k$. If there is not yet a proper item stored under $k$, then $M(k)$ should return an otherwise unused item $\bot$. By

$M[k \mapsto d]$ we denote the updated database $M'$, where $M'(j) = d$ if $j = k$, and $M'(j) = M(j)$ otherwise.

The two components PRE and RDB are specified as state machines (Fig. 5(a), 5(b)). We assume that there is a given function $f : Data \to Data$, that handles the preprocessing for a single datum.

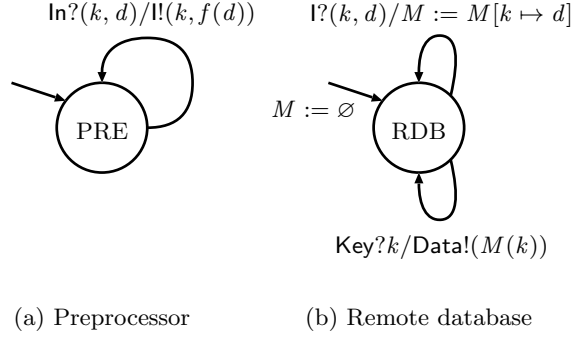

(a) Preprocessor    (b) Remote database

**Fig. 5.** Component specifications

In order to reduce the transmission time for the entries, we now want to transmit for each entry only the difference of the entry's data with respect to the already stored data for that key; the differences are assumed to be smaller in size than the data itself. Of course, the first entry for each key will need to be transmitted completely.

We are not interested in the algorithmic aspects of the computation of the difference between old data and new data; we just assume that the difference between two data items is itself an element of $Data$, and that there is a function

$$\Delta : Data \times Data \to Data$$

that computes the difference between old and new data. Another function

$$\rho : Data \times Data \to Data$$

reconstructs the new data given old data and the difference. We require that

$$\rho(d_{old}, \Delta(d_{old}, d_{new})) = d_{new} \quad (\dagger)$$

To simplify our specifications, we also assume that

$$\Delta(\bot, d) = d, \quad \rho(\bot, \delta) = \delta$$

These two functions can be extended to streams, where they take a database M as an additional parameter:

$$\Delta^*_M (\langle\rangle) = \langle\rangle$$
$$\Delta^*_M ((k, d) \,\&\, x) = (k, \Delta(M(k), d)) \,\&\, \Delta^*_{M[k \mapsto d]} (x)$$

$$\rho^*_M (\langle\rangle) = \langle\rangle$$
$$\rho^*_M ((k, \delta) \,\&\, x) = (k, \rho(M(k), \delta)) \,\&\, \rho^*_{M[k \mapsto \rho(M(k), \delta)]} (x)$$

Informally, the system modification is simple: the preprocessor is extended with a local database; for each new entry the difference to the old is computed and forwarded. The remote database reads the input, computes the new value out of the stored value and received difference, and stores this new value in its database. One possible design for this modification is to introduce encoding and decoding components, that compute the differences and reconstruct the original data, respectively.

In the sequel, we show how this refinement can be justified with our rule system. The modification consists of eight steps.

*Step 1: Adding components.* First, we introduce two new components to the system by two applications of the refinement rule. The new components, ENC and DEC, are not connected to any other component in the system.

After this refinement step, the system looks as follows:

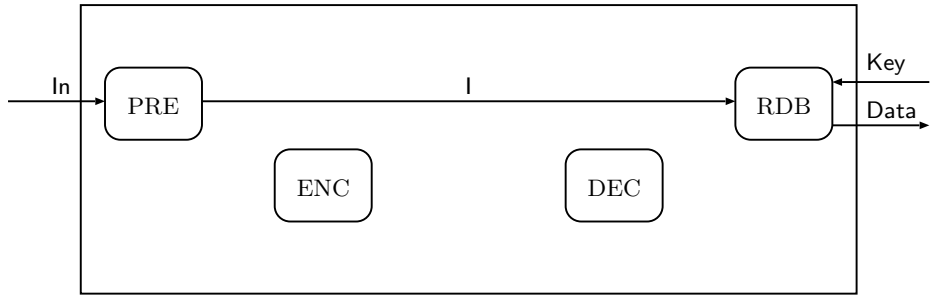

*Step 2: Adding output channels.* Now we add an output channel D to ENC, and an output channel R to DEC. Since these channels are neither part of the system interface, nor previously connected to any component, the premises of the refinement rule for the addition of channels are satisfied. Note that the contents of the channel are so far completely undefined, and the components ENC and DEC are therefore now nondeterministic. Nevertheless, the behavior of the system itself is unchanged, since the data on the new channels is unused throughout the system.

The following figure depicts the system after this refinement step:

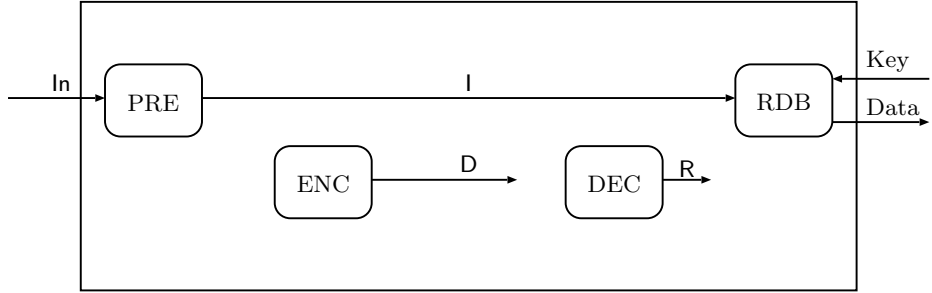

*Step 3: Adding input channels.* We now connect the channel I to the encoder ENC. The encoder still ignores the additional input, however, and hence the output D of ENC is still arbitrary. Similarly, we connect D to the decoder.

The system now looks like this:

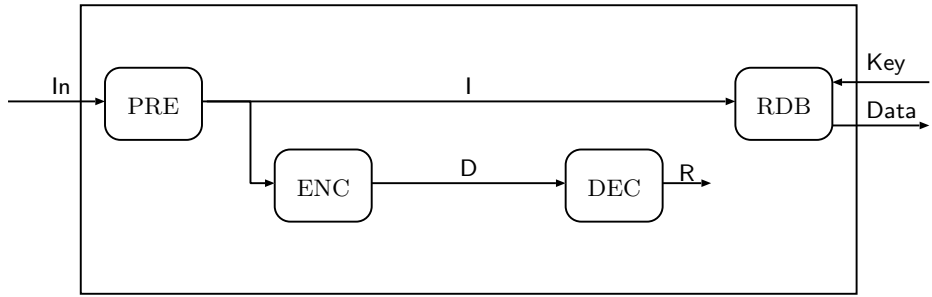

*Step 4: Behavioral refinement.* Now we constrain the channels D and R to carry the differences of the data on I and the reconstructed data, respectively. This is accomplished by restricting the behavior of ENC and DEC, and we can use the simple behavioral refinement rule for this step.

The encoder component is now specified as follows:

$(\text{ENC}, \{\mathsf{I}\}, \{\mathsf{D}\}, \beta)$

where

$\forall l : l \mid_{\{\mathsf{D}\}} \in \beta(l \mid_{\{\mathsf{I}\}}) \Leftrightarrow l(\mathsf{D}) = \Delta^*_\varnothing(l(\mathsf{I}))$

Thus, the encoder just applies the difference function $\Delta^*$ to its input stream I.

Similarly, we define the behavior of DEC as an application of the restoration function $\rho^*$. Since until now the behavior of the components was completely unspecified, this refinement is obviously correct.

The structure of the system remains unchanged.

*Step 5: Adding an input channel.* We now connect the channel R to the remote database. The behavior of RDB still ignores the additional input, however.

This step gives us the following system:

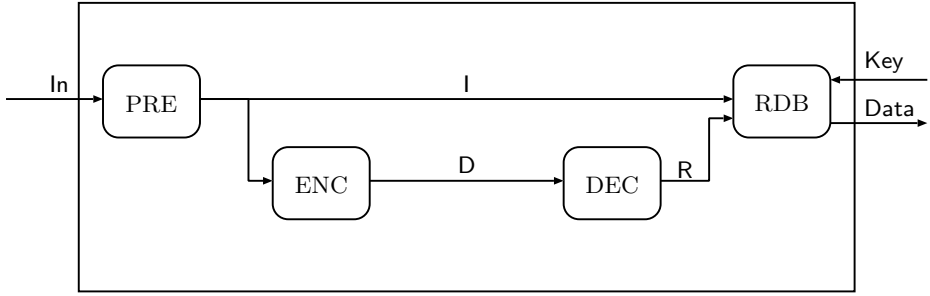

*Step 6: Behavioral refinement with invariant.* Now we want the remote database to store the data transmitted on R instead of that on I. Conversely, the input via I should be ignored.

The new behavior can again be specified as a state transition diagram; it looks just like the one in Fig. 5(b), except that now the upper transition reads from channel R and not from channel I.

Unfortunately, we cannot prove this refinement step with the simple behavioral refinement rule used in Step 4. The reason is that after the refinement the behavior of RDB is only then still correct, if the data on R is the same as that on I. Since neither R nor I is controlled by RDB, this cannot be proven locally.

The solution here is to use the behavioral refinement rule with an invariant. Intuitively, we know that encoding and then decoding the processed data from PRE yields the same data as that on I. We can formalize this knowledge with the following invariant:

$$\Psi(l) =_{def} l(\mathsf{I}) = \rho^*_\varnothing(\Delta^*_\varnothing(l(\mathsf{I})))$$

To show that $\Psi$ is indeed an invariant we prove the following stronger property, which implies $\Psi$:

$$\forall\, x, \forall\, M : \rho^*_M(\Delta^*_M(x)) = x$$

The proof is by induction on $x$:

- If $x = \langle\,\rangle$, we have for all $M$: $\Delta^*_M(x) = \langle\,\rangle$, and hence $\rho^*_M(\Delta^*_M(\langle\,\rangle)) = \langle\,\rangle$.
- If $x = (k, d)\,\&\,y$, then for an arbitrary $M$:

$$\begin{aligned}
&\rho^*_M(\Delta^*_M((k,d)\,\&\,y)) = \\
&\quad \rho^*_M((k, \Delta(M(k), d))\,\&\,\Delta^*_{M[k\mapsto d]}(y)) = \\
&\quad (k, \rho(M(k), \Delta(M(k), d)))\,\&\, \\
&\qquad \rho^*_{M[k\mapsto \rho(M(k),\Delta(M(k),d))]}(\Delta^*_{M[k\mapsto d]}(y)) = \\
&\quad (k, d)\,\&\,\rho^*_{M[k\mapsto d]}(\Delta^*_{M[k\mapsto d]}(y)) = \\
&\quad (k, d)\,\&\,y
\end{aligned}$$

The first two equalities follow from the definition of $\Delta^*$ and $\rho^*$, respectively; the third equality follows from the property (†) on page 14. The fourth equality uses the induction hypothesis.

It is straightforward to then prove the premises of the behavioral refinement rule with invariant for $\Psi$. The structure of the system remains unchanged.

*Step 7: Removing an input channel.* The new behavior of RDB now depends only on the data on R, and not on that in I. (An easy syntacical criterion is that RDB's state machine does not read from R any more.) Thus, we can disconnect I from RDB. The channel I now only feeds the encoder.

The new system looks as follows:

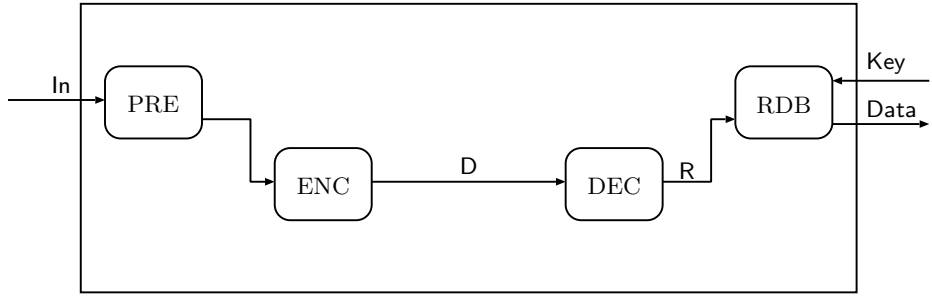

*Step 8: Folding subsystems.* In the last refinement step, we fold the two components PRE and ENC to a new component PRE′, and DEC together with RDB to a new component RDB′:

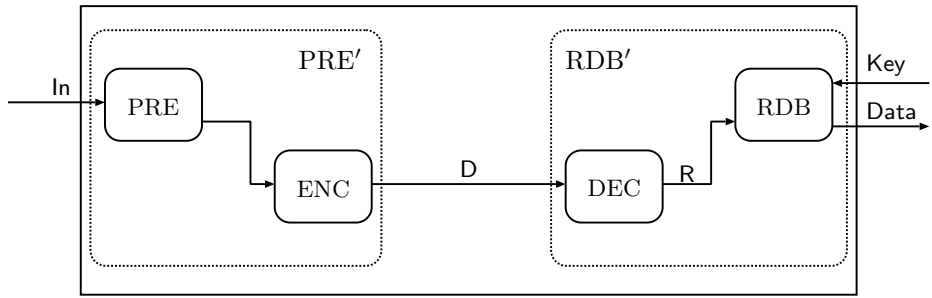

*Comments on the transformation.* The refinement steps described above are not fully formal; they cannot be, since we did not use a properly formalized description of the component behaviors. Of course, state transition diagrams can be given a mathematical semantics [3], and in [15] a refinement calculus for state transition diagrams is defined. Still, we hope that the example shows that although each individual refinement rule is quite simple, they can be used together for complex system transformations.

In fact, the example can be regarded as the derivation of a more complex refinement pattern: We now know how to transform systems as in our example, provided we have encoding and decoding functions $\Delta$ and $\rho$ that enjoy the property (†) on page 14.

As expected, the behavioral rule with invariant is the most complex rule to apply. In general, it is a difficult task for the system designer to find a proper invariant $\Psi$ that is both easy to establish and sufficiently strong to use. The maximal invariant $\Psi(l) = \text{True}$ leads to our initially given simple refinement rule without an invariant. The minimal possible $\Psi$ gives an exact description of the internal behavior of a system, but it is often difficult to find and too complex to use.

In practice, the invariant $\Psi$ often only affects a subset of the system channels, and typically forms an abstract environment of the component to modify. In our case, the invariant refers only to components introduced and removed in the refinement sequence. When we combine our rule applications to build the more complex rule mentioned above, we can therefore apply this new rule without having to prove the invariant again. We expect that there are many architectural refinement patterns, where invariants can be shown to hold once and for all, so that they can be applied in case tools without further proof obligations.

## 6   Conclusion

We believe that the question of how to manipulate and adapt an architecture during system development has not been adequately addressed so far. In particular, a precise calculus, dealing with simple addition and removing of channels and components in a data flow-based architectural style has—to our knowledge—not been considered before. This is somewhat surprising given the long history of data flow concepts in computer science [7, 18] and given the amount of work on components that partly also treat refinement [8, 16, 9, 1].

The most promising attempt at architecture refinement so far has been given in [13], where different software architectures are related by a kind of refinement mapping. Their approach, however, requires a "faithful implementation" of one architecture by the other, which makes the refinement definition somewhat complicated. In our approach, which supports underspecification, refinement is essentially just logical implication.

An interesting direction is the description of component behaviors by state machines as indicated in our example and the application of state machine refinement rules (as defined e.g. in [15]) for component behavior refinement. A concrete description technique for component behavior is essential for the proof of the invariant in the behavior refinement rule. For state machine refinement rules, too, underspecification is essential.

The simplicity and the compositionality of the calculus allow us to build a set of more powerful and more specific rules that describe design patterns as transformations, or embed newly defined data flow structures into a given architecture.

Of course refinement is not limited to pipe and filter architectures as used in this paper, but can be applied to a variety of styles, such as rule-based systems, interpreters, communicating systems or event based systems [16]. Finally, architecture refinement is by no means limited to software systems. A promising application area is hardware design and in particular the codesign of hardware and software components, where frequently a basic design has been changed because of cost or performance issues. Moreover, the simpler description techniques used in hardware design and the finite-state nature of such systems open the door to automatic verification of the refinement rule premises.

*Acknowledgement* This work originates from the SysLab project (supported by the DFG Leibniz program, by Siemens-Nixdorf and by Siemens Corporate Research), from FORSOFT (supported by the Bayerische Forschungsstiftung) and from the Sonderforschungsbereich 342 (supported by the DFG).